\documentclass[useAMS,usenatbib,onecolumn]{mn2e}
\usepackage{amsmath,graphicx,txfonts}
\newcommand{\mr}{\mathrm}

\def\C{\mathcal C}
\def\R{\mathcal R}

\newcommand{\be}{\begin{equation}}
\newcommand{\ee}{\end{equation}}

\newcommand{\beq}{\begin{eqnarray}}
\newcommand{\eeq}{\end{eqnarray}}
\newcommand{\ensav}[1]{\left\langle #1 \right\rangle}
\newcommand{\vt}{\vartheta}
\newcommand{\vp}{\varphi}

\newcommand{\bvt}{\boldsymbol\vt}
\newcommand{\G}{\Gamma}
\renewcommand{\d}{{\rm d}}
\newcommand{\vx}{\mathbf x}
\newcommand{\abs}[1]{| #1 |}

\def\aap{A\&A}
\def\apj{ApJ}
\def\apjl{ApJL}

\def\mnras{MNRAS}

\def\prd{Phys.~Rev.~D}        

\begin{document}
\title[The third-order ring statistics]{A new third-order cosmic shear statistics: Separating E/B-mode correlations on a finite interval}
\author[Krause, Schneider \& Eifler]{Elisabeth Krause$^1$,Peter Schneider$^2$ \& Tim Eifler$^3$
\\$^1$California Institute of Technology, MC 249-17, Pasadena, CA, 91125, USA, ekrause@tapir.caltech.edu
\\$^2$Argelander Institut f\"ur Astrophysik, Universit\"at Bonn, Auf dem H\"ugel 71, 53121, Bonn, Germany
\\$^3$Center for Cosmology and Astro-Particle Physics, The Ohio State University, 191 W. Woodruff Avenue, Columbus, OH 43210, USA
}
\date{}
\pagerange{\pageref{firstpage}--\pageref{lastpage}} \pubyear{2011}
\maketitle
\label{firstpage}
\begin{abstract}
Decomposing the shear signal into E and B-modes properly, i.e. without leakage of B-modes into the E-mode signal and vice versa, has been a long-standing problem in weak gravitational lensing. At the two-point level this problem was resolved by developing the so-called ring statistics, and later the COSEBIs; however, extending these concepts to the three-point level is far from trivial. Currently used methods to decompose three-point shear correlation functions (3PCFs) into E- and B-modes require knowledge of the 3PCF down to arbitrary small scales. This implies that the 3PCF needs to be modeled on scales smaller than the minimum separation of 2 galaxies and subsequently will be biased towards the model, or, in the absence of a model, the statistics is affected by E/B-mode leakage (or mixing). \\
In this paper we derive a new third-order E/B-mode statistic that performs the decomposition using the 3PCF only on a finite interval, and thereby is free of any E/B-mode leakage while at the same time relying solely on information from the data. In addition, we relate this third-order ring statistics to the convergence field, thereby enabling a fast and convenient calculation of this statistic from numerical simulations.
We note that our new statistics should be applicable to corresponding E/B-mode separation problems in the CMB polarization field.\end{abstract}
\begin{keywords}
cosmology: theory, large-scale structure of Universe\end{keywords}
\section{Introduction}
Cosmic shear, the distortion of light from distant galaxies by the tidal gravitational field of the intervening large-scale structure, is an excellent tool to probe 
the matter distribution in the universe. The statistics of the image distortions are related to the statistical properties of the large-scale matter distribution and the geometry of the universe, and 
can thereby be used to constrain cosmology. Current results already demonstrate the power of cosmic shear observations at constraining the clustering amplitude 
$\sigma_8$ and the matter density $\Omega_{\mr m}$ \citep[e.g.,][]{ Fu, Tim,Eric}. 
Furthermore, cosmic shear provides an ideal tool to study dark energy 
through measuring the growth of structure with large future surveys like KIDS\footnote{http://www.astro-wise.org/projects/KIDS}, DES\footnote{http://www.darkenergysurvey.org/}, LSST\footnote{http://www.lsst.org/lsst/} \citep{LSST}, or Euclid\footnote{http://sci.esa.int/euclid} \citep{Euclid}.
The large volume probed by these surveys will enable us to measure not only the power spectrum, but also higher-order statistics with unprecedented precision. As the evolved density field is non-Gaussian, the three-point correlation function and its Fourier space equivalent, the bispectrum, contain significant cosmological information complementary to the more commonly used two-point statistics and are a powerful tool for breaking parameter degeneracies \citep{Takada04}.\\
The upcoming weak lensing experiments will limit the 
statistical uncertainties to the percent level. In order to extract cosmological information from these cosmic shear experiments, the increased data quality needs to be accompanied by a thorough treatment of a wide range of systematic errors, from photometric redshifts and galaxy shape measurements to the removal of astrophysical contaminants.\\
If the shear estimated from observed galaxy shapes is solely caused by gravitational lensing, then it should consist only of a ``gradient component'', the so-called E-mode shear. B-modes (or curl components) cannot be generated by gravitational light deflection to leading order, and higher-order corrections are expected to be very small. Hence observing any B-mode pattern indicates remaining systematics in the shear analysis.\\
Decomposing the observed shear field directly into E/B-modes \citep[e.g][]{Bunn03} is complicated by the complex mask geometry of weak lensing observations.
At the two-point statistics level, an E/B-mode decomposition is commonly performed using the aperture mass dispersion \citep{Map2} and related measures \citep[e.g.][]{C02}, which can be calculated from the measured shear two-point correlation function (2PCF) and is thus not affected by the masking geometry. However, these methods assume that the 2PCF is known either from $\theta = 0$ to some finite angular value (aperture mass dispersion) or to arbitrarily large separations. However, in reality the 2PCF can only be measured on a finite interval $[\theta_{\mr{min}} , \theta_{\mr{max}}]$, where the lower boundary is caused by inability to measure the shape of image pairs with very small angular separation. As \citet{Kilbinger06} pointed out, lack of shear-correlation measurements on small scales leads to an underestimation of the aperture mass dispersion on small scales and causes an apparent mixing of E- and B-modes with this type of estimator. \citet{SK07, ring2} and \citet{COSEBI} develop statistical measures for an exact E/B-mode decomposition based on 2PCFs known only on a finite interval $[\theta_{\mr{min}} , \theta_{\mr{max}}]$.\\
At the three-point statistics level, \citet{JBJ04} and \citet{S05} introduced E/B-mode separating shear measures which assume knowledge of the 3PCF down to arbitrarily small scales. \citet{Shi11} derived a general condition for the E/B-mode decomposition of lensing three-point statistics, but the construction of filter functions with finite support based on this condition is far from straight forward.
In this paper we derive an extension of the 2PCF ring statistics \citep{SK07,ring2} to an exact E/B-mode decomposition of  shear three-point correlation functions on a finite interval.\\
In order to constrain cosmology with third-order shear statistics it is important to obtain the corresponding predictions from a large suite of cosmological numerical simulations in a reasonable time and with limited computational effort. We facilitate this by giving an expression of the third-order ring statistics in terms of the convergence field, thereby avoiding the time-consuming calculation of the shear 3PCF for each simulation.
\section{Shear three-point correlation functions}
\label{sec:3PCF}
\begin{figure}
\includegraphics[width =0.45\textwidth ]{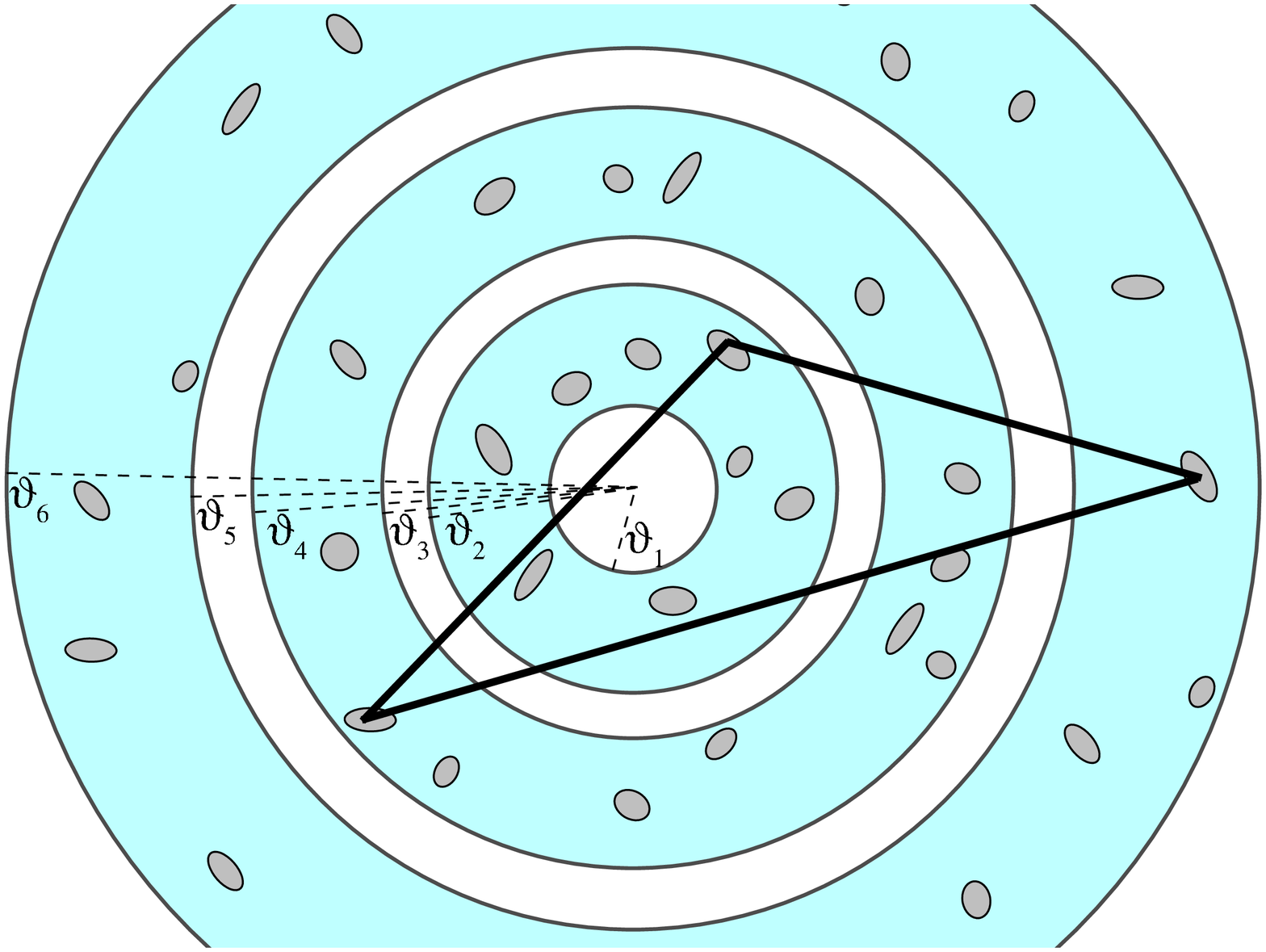}
\includegraphics[width = 0.55\textwidth]{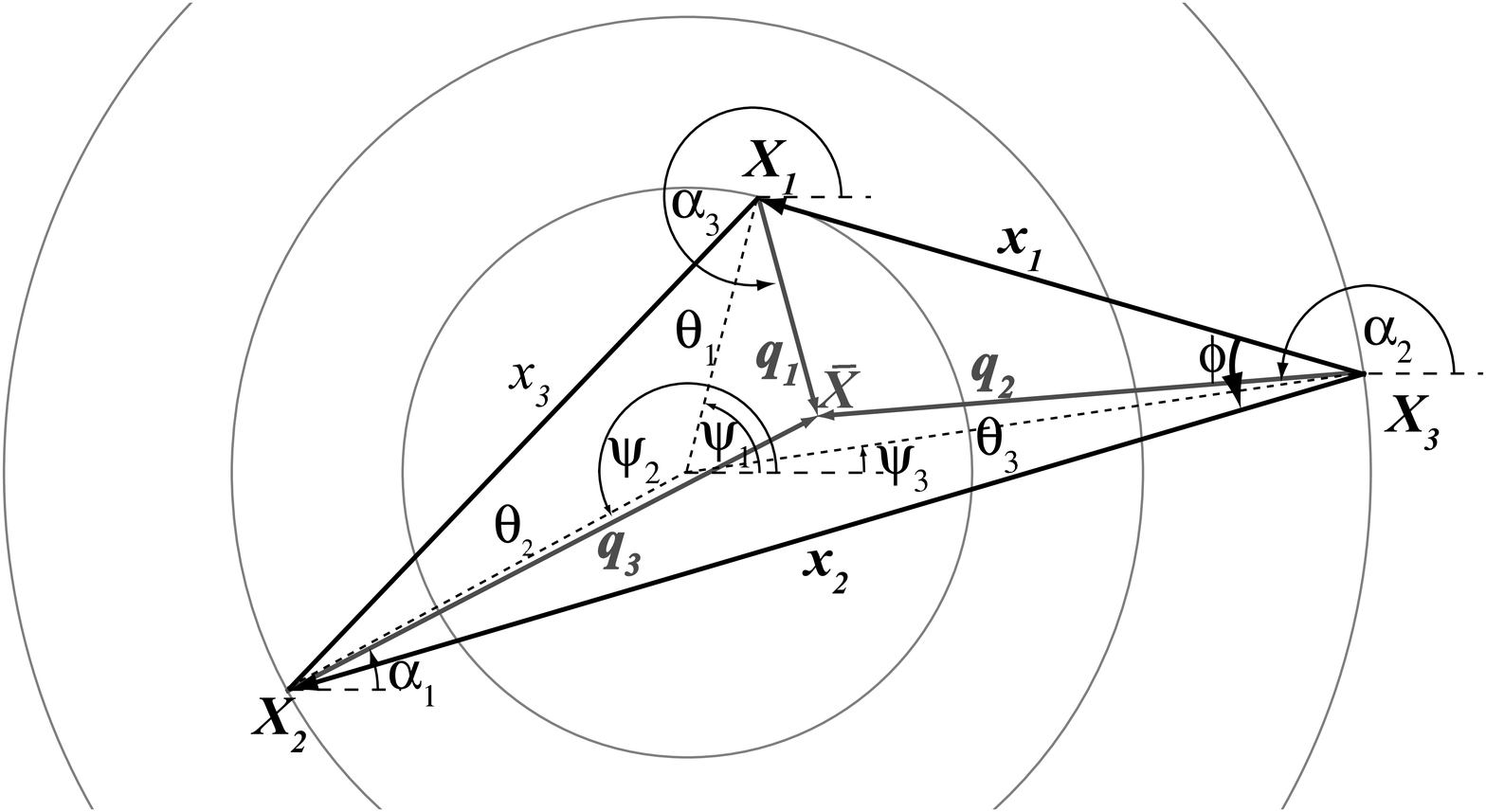}
\caption{\emph{Left:} Concept of the third-order ring statistics $\ensav{\cal R\cal R\cal R}\left(\vt_1,\vt_2,...,\vt_6\right)$. E/B-modes are separated on a finite interval by correlating the shear of galaxy triplets located within three concentric annuli, as illustrated by the thick black triangle. The minimum separation between galaxies in the above geometry is $\min(\vt_3-\vt_2, \vt_5-\vt_4)$ and the maximum separation is $\vt_6+\vt_4$.\newline
\emph{Right:} Geometry of a triangle in the third-order ring statistics. The triangle vertices $\mathbf X_j$ are located on three concentric circles of radius $\theta_j$ and have polar angles $\psi_j$. The triangle centroid is $\bar{\mathbf X}$. In the centroid projection the shear at each $\mathbf X_j$ is projected onto the centroid along direction $\mathbf q_j$, the line connecting $\mathbf X_j$ with the centroid. $\vp_j$ is the orientation angle of vector $\mathbf q_j$. Finally, $\phi$ is the inner angle of the triangle at $\mathbf X_3$ which we will use when parameterizing a triangle in terms of two side lengths $x_1, x_2$ and angle $\phi$.}
\label{fig:concept}
\label{fig:angles}
\end{figure}

We first introduce the shear three-point correlation function (3PCF): Consider a triangle in the complex plane with vertices $\mathbf X_i$ and let $\gamma_\mu(\mathbf X_i)$, $\mu = 1,2$ be the Cartesian components of the shear at point $\mathbf X_i$. Unless otherwise noted, we will assume that the triangle is oriented such that $\mathbf X_1$,  $\mathbf X_2$, $\mathbf X_3$ are ordered counterclockwise around the triangle. We define $\mathbf x_1= \mathbf X_1- \mathbf X_3$ and  $\mathbf x_2= \mathbf X_2- \mathbf X_3$ to be the sides of this triangle (c.f. Fig.~\ref{fig:angles}). We will use $\mathbf x_i$ to refer to complex numbers or vectors interchangeably, and denote their magnitude as $x_i$.\\
The Cartesian components of the shear 3PCF are defined as 
\be
\gamma_{\mu \nu \lambda}(\vx_1,\vx_2) \equiv \ensav{\gamma_\mu(\mathbf X_1)\gamma_\nu(\mathbf X_2)\gamma_\lambda(\mathbf X_3)}\,,
\ee
where we have assumed that the shear field is statistically homogeneous so that $\gamma_{\mu\nu\lambda}$ depends only on the side vectors $\vx_i$. Since one cannot form a tri-linear scalar from the product of three shears, the behavior of the Cartesian components of the shear 3PCF under rotations is complicated. In order to write the 3PCF in terms of tangential  ($\gamma_{\mr t})$  and cross components ($\gamma_{\times})$ of the shear which are parity eigenstates and have relatively simple transformation properties, one can project the complex Cartesian shear $\gamma^{\mr c} = \gamma_1 +\mr i \gamma_2$ into tangential and cross component  with respect to a chosen direction $\mathbf a_i$ with polar angle $\alpha_i$, 
\be
\gamma (\mathbf X_i; \alpha_i) \equiv \gamma_{\mr t} (\mathbf X_i; \alpha_i)+\mr i \gamma_{\times} (\mathbf X_i; \alpha_i) = -\left[\gamma_1(\mathbf X_i)+ \mr i \gamma_2(\mathbf X_i)\right]\mr e^{-2\mr i \alpha_i}
=- \gamma^{\mr c}(\mathbf X_i)\mr e^{-2\mr i \alpha_i} =  -\gamma^{\mr c}(\mathbf X_i)\,\mathbf a_i^{*2}/a_i^2\,.
\ee
If the directions of projection $\alpha_i$ are defined in terms of the vertices $\mathbf X_i$ and thus do not depend on an external coordinate system, then the tangential and cross shear are invariant under rotations of the triangle \citep{SL03, TJ03, ZS03}, and the 3PCF of these shear projections will only depend on the side \emph{lengths} $x_i$ and the orientation of the triangle (clockwise or counterclockwise). In the following we will use the centroid projection, where the shear at vertex $\mathbf X_i$ is projected along the direction $\mathbf q_i$ connecting $\mathbf X_i$ with the centroid $\bar{\mathbf X} = (\mathbf X_1+\mathbf X_2+\mathbf X_3)/3$, and $\alpha_i$ is the polar angle of this projection direction (see Fig.~\ref{fig:angles} for an illustration).\\
Following \citet{SL03} we define the (complex) natural components of the 3PCF which have relatively simple transformation properties
\beq
\nonumber \G^{(0)} (x_1,x_2,x_3) &\equiv& \ensav{\gamma \left(\mathbf X_1; \alpha_1\right)\gamma \left(\mathbf X_2; \alpha_2\right)\gamma \left(\mathbf X_3; \alpha_3\right)}=  -\ensav{\gamma^{\mr c} \left(\mathbf X_1\right)\gamma^{\mr c} \left(\mathbf X_2\right)\gamma^{\mr c} \left(\mathbf X_3\right)}\mr e^{-2\mr i (\alpha_1+\alpha_2+\alpha_3)}\\
\nonumber \G^{(1)} (x_1,x_2,x_3) &\equiv& \ensav{\gamma^* \left(\mathbf X_1; \alpha_1\right)\gamma \left(\mathbf X_2; \alpha_2\right)\gamma \left(\mathbf X_3; \alpha_3\right)}=  -\ensav{\gamma^{\mr c*} \left(\mathbf X_1\right)\gamma^{\mr c} \left(\mathbf X_2\right)\gamma^{\mr c} \left(\mathbf X_3\right)}\mr e^{-2\mr i (-\alpha_1+\alpha_2+\alpha_3)}\\\
\G^{(2)} (x_1,x_2,x_3) &\equiv& \ensav{\gamma \left(\mathbf X_1; \alpha_1\right)\gamma^* \left(\mathbf X_2; \alpha_2\right)\gamma \left(\mathbf X_3; \alpha_3\right)},\;\; \G^{(3)} (x_1,x_2,x_3) \equiv \ensav{\gamma \left(\mathbf X_1; \alpha_1\right)\gamma \left(\mathbf X_2; \alpha_2\right)\gamma^{*} \left(\mathbf X_3; \alpha_3\right)}\,.
\label{eq:Gamma}
\eeq
$\G^{(0)}$ is invariant under cyclic permutations of arguments; the other three components transform into each other: $\G^{(1)}(x_1,x_2,x_3) = \G^{(2)}(x_3,x_1,x_2) =\G^{(3)}(x_2,x_3,x_1)$, etc..
A different parameterization of oriented triangles is in terms of two sides and their inner angle, e.g. $x_1$,$x_2$, and $\phi$ (c.f. Fig.~\ref{fig:angles}). We choose the convention $\phi\in[-\pi,\pi]$, such that $\phi> 0$ corresponds to $\mathbf X_1$,  $\mathbf X_2$, $\mathbf X_3$  being ordered counter clock wise (''positive orientation") and $\phi< 0$ corresponds to clock wise ordering (''negative orientation''). 
\section{E/B-Mode separation}
To construct integrals which separate third-order E- and B-mode correlations we start from the circle statistics $\C(\theta)$ \citep{C02,S02} which geometrically separates E- and B-modes by measuring the mean tangential and cross component of the shear on a circle of radius $\theta$ around the origin 
\be
 \C(\theta) = \C_{\mr t} (\theta) + \mr i \C_{\times}(\theta) = \frac{1}{2 \pi}\int_0^{2\pi} \d \psi\; \left(\gamma_{\mr t} + \mr i \gamma_{\times}\right)\left(\theta,\psi;\psi\right)
= -\frac{1}{2 \pi}\int_0^{2\pi} \d \psi\; \mr e ^{-2\mr i \psi}\gamma^{\mr c}\left(\theta,\psi\right)\;,
\ee
where $\psi$ is the polar angle on the circle, and in the last step we have rotated the tangential and radial shear into the cartesian components.
Following \citet{SK07} we now consider the shear inside an annulus $\vt_1\leq \theta \leq \vt_2$ and define the ring statistics $\R$
\be
\R = \R_{\mr t}  + \mr i \R_{\times}  = \int_{\vt_1}^{\vt_2} \d \theta\, W(\theta;\vt_1,\vt_2)\,\C(\theta)\,,
\label{eq:Rdef}
\ee
which is a function of two radii $\vt_i$ and $\vt_j$, and where $W(\theta;\vt_i,\vt_j)$ is a normalized weight function
\be
\int_{\vt_i}^{\vt_j} \d \theta\; W(\theta;\vt_i,\vt_j) =1\,,
\ee
and $W = 0$ outside the annulus, i.e. if $\theta < \vt_i$ or $\theta>\vt_j$.
From this definition we construct the third-order ring statistics as the correlation of the weighted mean shear in three concentric annuli 
with radii $\vt_1 \leq \theta_1,\leq \vt_2 <\vt_3\leq \theta_2\leq \vt_4 <\vt_5\leq \theta_3\leq \vt_6$ (cf. Fig.~\ref{fig:angles}),
\beq
\ensav{\R\R\R}(\boldsymbol\vt) &=&  \int_{\vt_1}^{\vt_2} \d \theta_1 W(\theta_1;\vt_1,\vt_2) \int_{\vt_3}^{\vt_4} \d \theta_2 W(\theta_2;\vt_3,\vt_4) \int_{\vt_5}^{\vt_6} \d \theta_3 W(\theta_3;\vt_5,\vt_6)\;\ensav{\C(\theta_1)\,\C(\theta_2)\,\C(\theta_3)}
 \label{eq:CCCgamma} \\
\ensav{\R^*\R\R}(\boldsymbol\vt)  &=&  \int_{\vt_1}^{\vt_2} \d \theta_1 W(\theta_1;\vt_1,\vt_2) \int_{\vt_3}^{\vt_4} \d \theta_2 W(\theta_2;\vt_3,\vt_4) \int_{\vt_5}^{\vt_6} \d \theta_3 W(\theta_3;\vt_5,\vt_6)\;\ensav{\C^*(\theta_1)\,\C(\theta_2)\,\C(\theta_3)}
\label{eq:C*CCgamma} 
 \eeq
where we have used $\boldsymbol\vt = (\vt_1,...,\vt_6)$ to denote a six-tuple of radii.
Expanding these correlators in terms of the mean tangential and cross shear yields
\beq
\label{eq:C3parts} \ensav{\R\R\R}(\bvt)&=&  \left[\ensav{\R_{\mr t}\R_{\mr t}\R_{\mr t}}-\ensav{\R_{\times}\R_{\times}\R_{\mr t}}-\ensav{\R_{\times}\R_{\mr t}\R_{\times}}-\ensav{\R_{\mr t}\R_{\times}\R_{\times}}
+\mr i \left(-\ensav{\R_{\times}\R_{\times}\R_{\times}}+\ensav{\R_{\times}\R_{\mr t}\R_{\mr t}}+\ensav{\R_{\mr t}\R_{\times}\R_{\mr t}}+\ensav{\R_{\mr t}\R_{\mr t}\R_{\times}}\right)\right]\left(\bvt\right)\\
 \ensav{\R^*\R\R}(\bvt)&=&  \left[\ensav{\R_{\mr t}\R_{\mr t}\R_{\mr t}}+\ensav{\R_{\times}\R_{\times}\R_{\mr t}}+\ensav{\R_{\times}\R_{\mr t}\R_{\times}}-\ensav{\R_{\mr t}\R_{\times}\R_{\times}}
+\mr i \left(\ensav{\R_{\times}\R_{\times}\R_{\times}}-\ensav{\R_{\times}\R_{\mr t}\R_{\mr t}}+\ensav{\R_{\mr t}\R_{\times}\R_{\mr t}}+\ensav{\R_{\mr t}\R_{\mr t}\R_{\times}}\right)\right]\left(\bvt\right)
\label{eq:C3*parts}
\eeq
Note that the imaginary parts of (\ref{eq:C3parts}, \ref{eq:C3*parts}) vanish in the absence of parity-violating modes.\\
We analogously define the correlators $\ensav{\R\R^*\R}$ and $\ensav{\R\R\R^*}$ and separate E- and B- modes via
\beq
\label{eq:CEEE}\ensav{\R_{\mr E}^3}\left(\bvt\right)&=&\frac{1}{4}\mr{Re} \left[\ensav{\R\R\R}+\ensav{\R^*\R\R}+\ensav{\R\R^*\R}+\ensav{\R\R\R^*}\right](\bvt)\\
\label{eq:CEBB} \ensav{\R_{\mr E}\R_{\mr B}^2}\left(\bvt\right)&=&\frac{1}{4}\mr{Re} \left[-3\ensav{\R\R\R}+\ensav{\R^*\R\R}+\ensav{\R\R^*\R}+\ensav{\R\R\R^*}\right](\bvt)\\
\label{eq:CEEB}  \ensav{\R_{\mr E}^2\R_{\mr B}}\left(\bvt\right)&=&\frac{1}{4}\mr{Im} \left[3\ensav{\R\R\R}+\ensav{\R^*\R\R}+\ensav{\R\R^*\R}+\ensav{\R\R\R^*}\right](\bvt)\\\
\label{eq:CBBB} \ensav{\R_{\mr B}^3}\left(\bvt\right)&=&\frac{1}{4}\mr{Im} \left[-\ensav{\R\R\R}+\ensav{\R^*\R\R}+\ensav{\R\R^*\R}+\ensav{\R\R\R^*}\right](\bvt)\, ,
\eeq
where Eq.~(\ref{eq:CEEE}) corresponds to pure E-mode correlations, and Eq.~(\ref{eq:CBBB}) to parity violating third-order B-mode correlations. Equation~(\ref{eq:CEBB}) is a parity invariant correlation between E- and B-modes, and Eq.~(\ref{eq:CEEB}) is a parity violating correlation between E- and B-modes.\\
For brevity, the mixed terms (\ref{eq:CEBB},\ref{eq:CEEB}) are generalized expressions which are sensitive to B-modes in any of the annuli, i.e $ \ensav{\R_{\mr E}\R_{\mr B}^2}\left(\bvt\right) = (\ensav{R_{\mr t} R_\times R_\times} + \ensav{R_{\times} R_{\mr t} R_\times}+\ensav{R_\times R_\times R_{\mr t}})/3$, etc.. Instead one can also consider more localized B-mode measures like
\beq
\ensav{R_{\times} R_\times R_{\mr t}}\left(\bvt\right)&=&\frac{1}{4}\mr{Re} \left[\ensav{-\R\R\R}+\ensav{\R^*\R\R}+\ensav{\R\R^*\R}-\ensav{\R\R\R^*}\right](\bvt)\,,
\label{eq:CBBE}
 \eeq
which picks up correlations with B modes in the innermost and middle annulus, but is insensitive to B-modes in the outer annulus.
\section{Third-order ring statistics}
In this section we derive computationally advantageous expressions for the third-order ring statistics in terms of the shear 3PCF, and show their relation to the convergence bispectrum.
\subsection{Relation to the shear three-point functions}
We rewrite the third-order ring statistics in terms of the shear 3PCF by starting from the definition Eq.~(\ref{eq:CCCgamma})
\beq
\nonumber \ensav{\R\R\R}(\boldsymbol\vt) &=&  -\int_{\vt_1}^{\vt_2} \d \theta_1 W(\theta_1;\vt_1,\vt_2) \int_{\vt_3}^{\vt_4} \d \theta_2 W(\theta_2;\vt_3,\vt_4) \int_{\vt_5}^{\vt_6} \d \theta_3 W(\theta_3;\vt_5,\vt_6)\\
 \label{eq:RRRgamma} 
&&\times  \int_0^{2\pi} \frac{\d \psi_3}{2 \pi} \int_0^{2\pi} \frac{\d \psi_2}{2 \pi} \int_0^{2\pi} \frac{\d \psi_1}{2 \pi}\; \mr e^{-2\mr i (\psi_1+\psi_2+\psi_3)} \ensav{\gamma^{\mr c}\left(\theta_1,\psi_1\right )\gamma^{\mr c}\left(\theta_2,\psi_2\right)\gamma^{\mr c}\left(\theta_3,\psi_3\right)}\,.
\eeq
Noting that $\mathbf X_j = \theta_j \exp(\mr i\psi_j)$ and using Eq.~(\ref{eq:Gamma}), this can be rewritten as
\beq
\nonumber \ensav{\cal{RRR}}\left(\bvt\right)&=&\int\frac{\d^2 X_1}{2\pi\, | \mathbf X_1|} \;W(| \mathbf X_1|;\vt_1,\vt_2)\int\frac{\d^2 X_2}{2\pi\, | \mathbf X_2|}\;W(|\mathbf X_2|;\vt_3,\vt_4)
\int\frac{\d^2 X_3}{2\pi\, | \mathbf X_3|}\;W(|\mathbf X_3|;\vt_5,\vt_6)\\
&&\times\exp(2{\rm i}(\alpha_1+\alpha_2+\alpha_3-\psi_1-\psi_2-\psi_3))\;
\Gamma^{(0)}(\mathbf X_1-\mathbf X_3,\mathbf X_2-\mathbf X_3)\;,
\label{eq:RGamma}
\eeq
where $\Gamma^{(0)}$ is the shear 3PCF measured relative to the
centroid, so that the $\alpha_i$ are the directions of the point $\mathbf X_i$
to the centroid $\mathbf{\bar X}=(\mathbf X_1+\mathbf X_2+\mathbf X_3)/3$. Owing to circular symmetry,
we can set $\psi_3=0$; equivalently, one can use relative polar angles
$\Delta \psi_j=\psi_j-\psi_3$ and show that the integrand depends only
on these relative angles.\\
As $\Gamma$ is measured within discrete angular bins, while the weight functions and geometric factors in Eq.~(\ref{eq:RGamma}) can be evaluated continuously, it is numerically more stable to rewrite the third-order ring statistics such that only the three outermost integrals contain the shear 3PCF and the inner integrals can be evaluated numerically to arbitrary precision. With $\mathbf x_j=\mathbf X_j-\mathbf X_3=\theta_j{\rm e}^{{\rm i}\psi_j}-\theta_3$ for $j=1,2$,
\beq
\nonumber \ensav{\cal{RRR}}\left(\bvt\right)&=&\frac{1}{(2\pi)^2} \int \d^2 x_1\int \d^2 x_2\,\Gamma^{(0)}(\mathbf x_1,\mathbf x_2)
\int \d \theta_3\;W(\theta_3;\vt_5,\vt_6)\,W(|\mathbf x_1+\boldsymbol \theta_3|;\vt_1,\vt_2)
\,W(|\mathbf x_2+\boldsymbol \theta_3|;\vt_3,\vt_4)\\
&&\times\frac{1}{|\mathbf x_1+\boldsymbol \theta_3||\mathbf x_2+\boldsymbol \theta_3|}\;\exp(2{\rm i}(\alpha_1+\alpha_2+\alpha_3-\psi_1-\psi_2) \;,
\eeq
where we have used $\boldsymbol \theta_3$ to denote a complex number with zero imaginary part for consistency. We have ${\rm e}^{2{\rm i}\alpha_j}=\mathbf q_j/\mathbf q_j^*$, with
$\mathbf q_1=(2\mathbf x_1-\mathbf x_2)/3$, $\mathbf q_2=(2\mathbf x_2-\mathbf x_1)/3$, $\mathbf q_3=-(\mathbf x_1+\mathbf x_2)/3$, and ${\rm e}^{{\rm i}\psi_j}=(\mathbf x_j+\boldsymbol \theta_3)/\theta_j$, so that
\be
{\rm e}^{-2{\rm i}\psi_j}={\mathbf x_j^*+\boldsymbol \theta_3^* \over \mathbf x_j+\boldsymbol \theta_3}\;.
\ee
Thus,
\beq
\nonumber \ensav{\cal{RRR}}\left(\bvt\right)&=&\frac{1}{(2\pi)^2} \int \d^2 x_1\int \d^2
x_2\,\Gamma^{(0)}(\mathbf x_1,\mathbf x_2)
\;\int \d \theta_3\;W(\theta_3;\vt_5,\vt_6)\,W(|x_1+\boldsymbol \theta_3|;\vt_1,\vt_2)
\,W(|x_2+\boldsymbol \theta_3|;\vt_3,\vt_4)\;\\
 && \times\frac{1}{|\mathbf x_1+\boldsymbol \theta_3||\mathbf x_2+\boldsymbol \theta_3|}\;\
{\mathbf q_1 \mathbf q_2 \mathbf q_3\over \mathbf q_1^* \mathbf q_2^* \mathbf q_3^*}\,
{\mathbf x_1^*+\boldsymbol \theta_3^* \over \mathbf x_1+\boldsymbol \theta_3}\,
{\mathbf x_2^*+\boldsymbol \theta_3^* \over \mathbf x_2+\boldsymbol \theta_3} \;.
\eeq
Finally, if $\vp_i$ is the polar angle of $\mathbf x_i$, and
$\phi=\vp_2-\vp_1$ is the angle between $\mathbf x_2$ and $\mathbf x_1$, we obtain
\beq
\nonumber \ensav{\cal{RRR}}\left(\bvt\right)&=&\frac{1}{(2\pi)^2}
\int\d x_1\; x_1 \int\d x_2\;x_2 \int\d\phi\;
\Gamma^{(0)}(x_1,x_2,\phi) 
\\
\nonumber&&\times\;\int \d \theta_3\,W(\theta_3;\vt_5,\vt_6)\int\d \vp_1\,  W(\abs{x_1 {\rm e}^{{\rm i}\vp_1}+\theta_3};\vt_1,\vt_2)
\, W(\abs{x_2 {\rm e}^{{\rm i}(\vp_1+\phi)}+\theta_3};\vt_3,\vt_4)\;
\frac{1}{|\mathbf x_1+\boldsymbol \theta_3||\mathbf x_2+\boldsymbol \theta_3|}\;{\mathbf q_1 \mathbf q_2 \mathbf q_3\over \mathbf q_1^* \mathbf q_2^* \mathbf q_3^*}{\mathbf x_1^*+\theta_3 \over \mathbf x_1+\theta_3}\,
{\mathbf x_2^*+\theta_3 \over \mathbf x_2+\theta_3}\\
&\equiv& \frac{1}{(2\pi)^2}
\int\d x_1\; x_1 \int\d x_2\;x_2 \int\d\phi\;
\Gamma^{(0)}(x_1,x_2,\phi)\; \mathbf Z_0(x_1,x_2,\phi,\bvt)\,,
\label{eq:RRRfinal}
\eeq
where we have defined the complex filter function $\mathbf Z_0$ of the ring statistics in the last step.
\begin{figure}
\includegraphics[width =\textwidth ]{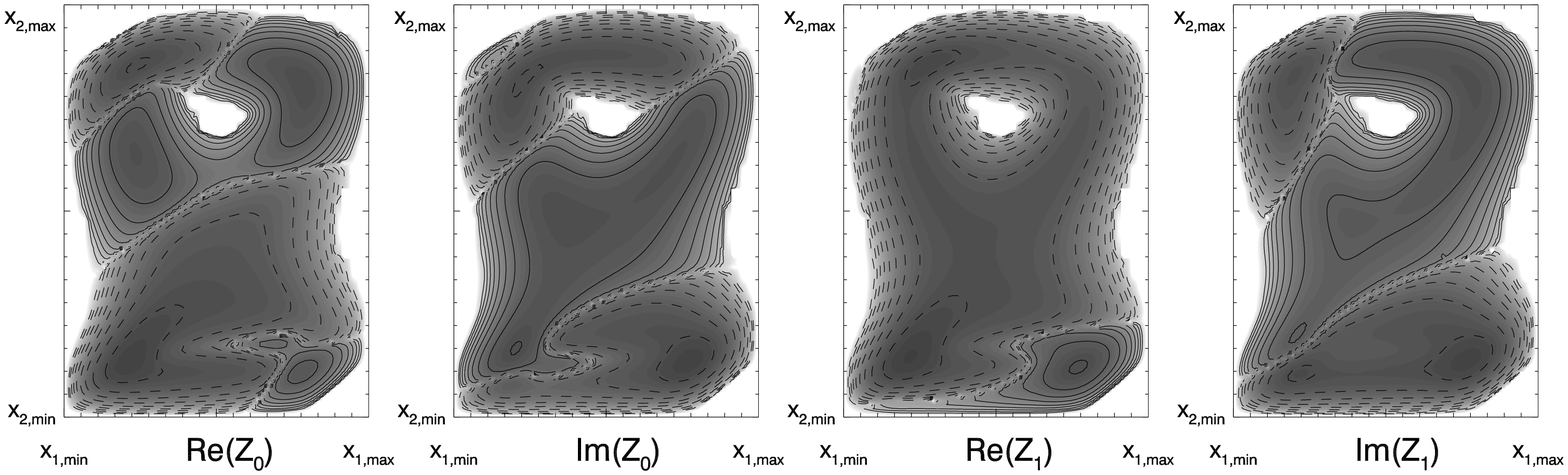}
\includegraphics[width =\textwidth ]{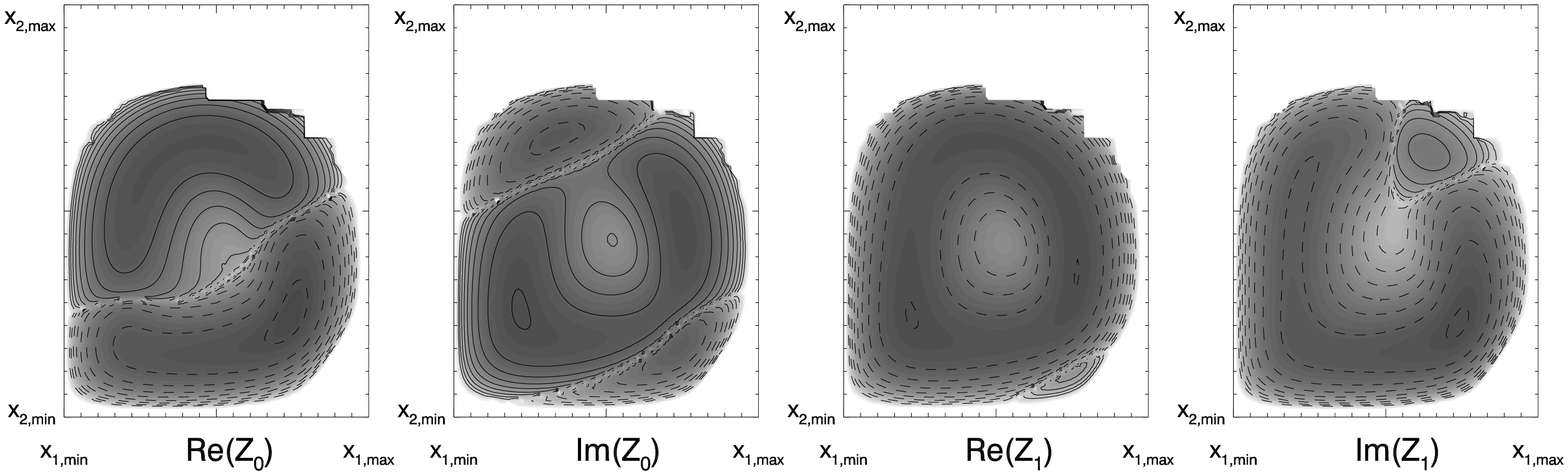}
\caption{Contours of the filter functions $\mathbf Z_{0,1}$ as a function of angular scales $x_1\in [\vt_5-\vt_2,\vt_6+\vt_2]$ and $x_2\in [\vt_5-\vt_4,\vt_6+\vt_4]$ for $\boldsymbol \vt = (1',2',3',4',5',6')$ with $\phi=\pi/8$ (top row) or $\phi=\pi/4$ (bottom row). Contour lines are evenly spaced with separation 0.5 dex ranging from $\abs{Z} =10^{-5}$ to $\abs{Z} =10^{-1.5}$, dashed lines indicate regions where $Z$ is negative. The filter functions vanish if the triangle configuration $(x_1,x_2,\phi)$ is not allowed in the ring statistics geometry (c.f. Fig~\ref{fig:concept}).}
\label{fig:Z}
\end{figure}
Note that the ratio of the $\mathbf q$'s does not depend on $\theta_3$ and
thus the evaluation of the filter function can be further simplified by reversing the-order of integration and moving this factor to the outer ($\vp_1$-) integral.\\
Expressions for the other correlations required for E/B-mode separation, which contain a complex conjugate ring statistic $\R^*$, are derived analogously. For the correlation involving the complex conjugate shear at vertex $\mathbf X_j$, the resulting expression analogous to Eq.~(\ref{eq:RRRfinal}) contains $\Gamma^{(j)}$ instead of $\Gamma^{(0)}$, $\mathbf q_j/\mathbf q^*_j$ is replaced by its complex conjugate (corresponding to $\alpha_j \rightarrow -\alpha_j$ in Eq.~(\ref{eq:Gamma})), and for $j = 1,2$ the factor $\mathbf x_j/\mathbf x^*_j$ is also replaced by its complex conjugate (corresponding to $\psi_j \rightarrow -\psi_j$ in the equivalent of Eq.~(\ref{eq:RRRgamma})), e.g.
\beq
\ensav{\R^*\R\R}\left(\bvt\right) &=&\frac{1}{(2 \pi)^2} \int\d x_1\; x_1 \int\d x_2\;x_2 \int\d\phi\;
\Gamma^{(1)}(x_1,x_2,\phi)\; \mathbf Z_1(x_1,x_2,\phi,\bvt)\,,
\label{eq:RRR2final}
\eeq
with
\beq
\mathbf Z_1(x_1,x_2,\phi,\bvt) \!\!\!&=& \!\!\!\!\!\int \d \theta_3\,W(\theta_3;\vt_5,\vt_6)\int\d \vp_1\,  W(\abs{x_1 {\rm e}^{{\rm i}\vp_1}+\theta_3};\vt_1,\vt_2)
\, W(\abs{x_2 {\rm e}^{{\rm i}(\vp_1+\phi)}+\theta_3};\vt_3,\vt_4)\;\frac{1}{|\mathbf x_1+\boldsymbol \theta_3||\mathbf x_2+\boldsymbol \theta_3|}
\frac{\mathbf q_1^* \mathbf q_2 \mathbf q_3}{\mathbf q_1 \mathbf q_2^* \mathbf q_3^*}
\frac{\mathbf x_1+\theta_3}{\mathbf x_1^*+\theta_3}\,
\frac{\mathbf x_2^*+\theta_3} {\mathbf x_2+\theta_3}\,.
\label{eq:Z1final}
\eeq
In the computation of the ring statistics one can choose any (normalized) radial weight function $W$ that fulfills $W(0;\vt_1,\vt_2)= 0$ even if $\vt_1=0$ (as the separation in tangential/cross shear is ill-defined on circle of radius $\theta=0$). To be specific, we choose
\be
W(\theta;\vt_i,\vt_j) = 30(\theta-\vt_i)^2(\vt_j-\theta)^2/(\vt_j-\vt_i)^5\,,
\ee
as in the computation of the second-order ring statistics \citep{SK07}. The shape of the third-order ring statistics filter functions $Z_{0,1}$ based on this choice for $W$ is illustrated in Fig.~\ref{fig:Z}.

\subsection{Relation to the bispectrum}
\label{sec:predictions}
\begin{figure}
\includegraphics[width =\textwidth ]{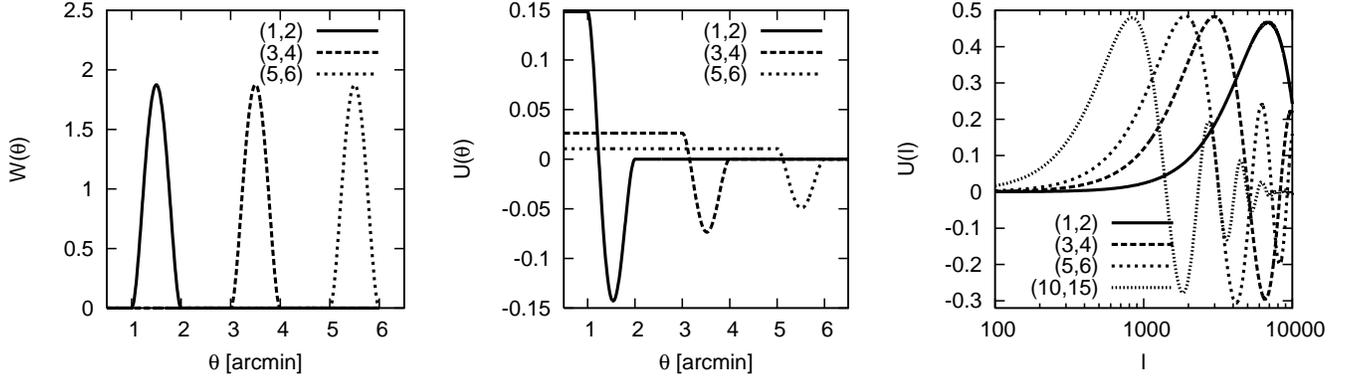}
\caption{Filter functions associated with $\R$ for different ring radii $(\vt_i,\vt_j)$. The left panel shows the radial filter function of the ring statistics, $W(\theta;\vt_i,\vt_j)$; the corresponding compensated aperture mass filter function $U_{\vt_i,\vt_j}(\theta)$ is shown in the middle panel, and the right panel illustrates the corresponding Fourier transform $\tilde{U}_{\vt_i,\vt_j}(l)$.} 
\label{fig:filter}
\end{figure}

In order to rewrite the third-order ring statistics in terms of the bispectrum we first relate it to the lensing convergence field $\kappa$, which is easier to express in terms of the convergence bispectrum than the shear 3PCF \citep[see][for details]{S05} as it contains fewer oscillatory phase factors. Expressing the ring statistics in terms of the convergence field also speeds up the measurement of $\ensav{\R_{\mr E}^3}$ in simulations considerably, as described below.\newline
Consider the convergence field smoothed with a radially symmetric filter $U_\vt(\theta)$ with characteristic scale $\vt$. If $U_\vt(\theta)$ is a compensated filter $\int \d\theta\;\theta\, U_{\vt}(\theta) = 0$, this convolution can be expressed in terms of the shear field as
\be
 \int\d^2\theta'\; U_{\vt}(\abs{\boldsymbol\theta'})\;\kappa(\boldsymbol\theta') = \int \d^2\theta\; Q_{\vt}(\abs{\boldsymbol\theta'})\;\gamma_{\mr t}(\boldsymbol\theta')\,,
\label{eq:Map}
\ee
where $U$ and $Q$ are related by \citep{Kaiser95, Schneider96, SK07}
\be
Q_\vt(\theta) = \frac{2}{\theta^2}\int_0^\theta \d \theta' U_\vt(\theta') - U_\vt(\theta)\;\;\;{\rm and}\;\;\; U_\vt(\theta) = \int_\theta^{\infty}\frac{2 \d \theta'}{\theta'} Q_\vt(\theta') - Q_\vt(\theta)\,.
\label{eq:QU}
\ee
As shown in \citet{SK07} the definition of the ring statistics $\R$ (Eq.~(\ref{eq:Rdef})) is equivalent to an aperture mass  $M_{\mr{ap}}(\vt_i,\vt_j)$ with two characteristic scales if 
\be
Q_{\vt_i,\vt_j}(\theta) = \frac{W(\theta;\vt_i,\vt_j)}{2\pi\,\theta\,}\,.
\label{eq:QW}
\ee
As the relation between the filter $Q$ and $U$ does not depend on the shape of $Q$, we can calculate the corresponding compensated filter $U_{\vt_i,\vt_j}(\theta)$ as in Eq.~(\ref{eq:QU}). The left and middle panel of Fig.~\ref{fig:filter} show the ring statistics filter $W(\theta;\vt_i,\vt_j)$ and the corresponding aperture mass filter $U_{\vt_i,\vt_j}(\theta)$ for different choices of ring radii $(\vt_i,\vt_j)$. As expected from Eq.~(\ref{eq:QU}), $U$ is constant for $\theta < \vt_i$, then becomes negative, and is zero for $\theta>\vt_j$.\\
Based on Eqs.~(\ref{eq:Map},\ref{eq:QU},\ref{eq:QW}), the third-order ring statistics of a pure E-mode field can be computed directly from simulated convergence maps by convolving the convergence field with different filters $U_{\vt_i,\vt_j}$ and correlating three filtered maps. With this approach one does not need to calculate the shear 3PCF, which are computationally expensive \citep[e.g.][]{JBJ04}.\\
Expressing $\R_\mr{E}$ as the convolution of $\kappa$ and $U_{\vt_i,\vt_j}$ also enables us to write down he third-order ring statistics of a pure E-mode field in terms of the convergence bispectrum $B_\kappa (l_1,l_2,l_3)$ \citep[c.f.][]{S05}
\be
\ensav{\R_{\mr E}^3}\left(\bvt\right) = 
\frac{1}{(2\pi)^3}\!\!\int\! \d l_1\, l_1\!\!\int\! \d l_2\, l_2 \!\!\int\! \d \phi\; B_\kappa(l_1,l_2,\sqrt{l_1^2+l_2^2-2 l_1 l_2 \cos{\phi}})
\tilde U_{\vt_1,\vt_2}(l_1)\tilde U_{\vt_3,\vt_4}(l_2)\tilde U_{\vt_5,\vt_6}( \sqrt{l_1^2+l_2^2-2 l_1 l_2 \cos{\phi}}) 
\,,
\label{eq:Bkappa}
\ee
with the Fourier transformed filter function $\tilde U(l) = \int \d \theta\, \theta \,\mr{J}_0(l\theta) U(\theta)$. The bispectrum filter functions for the third-order ring statistics are illustrated in the right panel of Fig.~\ref{fig:filter}.
\section{Conclusion}
Upcoming lensing surveys will provide data of unprecedented quality and enable us to conduct robust measurements of cosmic shear beyond the two-point level. These higher-order statistics contribute substantial information to cosmological constraints by breaking parameter degeneracies when combined with second-order shear statistics. Furthermore, three-pt statistics have the potential to improve our understanding of systematics effects in the data, e.g. a detection of third-order B-modes can be an additional indicator for unsolved problems in the data analysis.\\
When extracting third-order information from a high-quality data set it is therefore essential to use robust and unbiased theoretical methods that meet the quality of the data. We have introduced the third-order ring statistics, which separates the shear 3PCF into third-order E/B- mode correlations on a finite interval $[\theta_{\mr{min}},\theta_{\mr{max}}]$. Hence this statistic does not require knowledge of the 3PCF down to zero lag, where it is impossible to measure. Thus, unlike the third-order aperture mass statistics, it is not affected by apparent E/B-mode mixing \citep{Kilbinger06}.\\
Our main results are Eqs.~(\ref{eq:RRRfinal}, \ref{eq:RRR2final}, \ref{eq:Z1final}), which give compact expressions for the third-order ring statistics in terms of the shear 3PCF. Furthermore, in Sect.~\ref{sec:predictions} we give convenient expressions for computing the E-mode ring statistics from numerical simulations, and from the convergence bispectrum which facilitate the comparison with theoretical models for weak lensing three-point statistics \citep[e.g.][]{Valageas11}.\\
In addition to the cosmological information contained in the E-mode signal, our expression for third-order B-mode correlations opens a new window to detect remaining systematics in the data. For example, the various permutations of $\ensav{\R_{\mr E}\R_{\mr B}\R_{\mr E}}$ allow for an association of B-modes with a specific angular scales.\\
For the analysis of future shear 3PCF measurements, we recommend using Eq.~(\ref{eq:CEEE}) to obtain a clean third-order E-mode signal, and Eqs.~({\ref{eq:CEBB}, \ref{eq:CBBE}}) to test for remaining B-mode correlations.
\section*{Acknowledgements}
We thank Bhuvnesh Jain, Chris Hirata, and Mike Jarvis for useful discussions.\\
This work was supported by the Deutsche Forschungsgemeinschaft under
the Transregional Collaborative Research Center TR-33 `The Dark
Universe'.
EK is supported by the US National Science Foundation (AST-0807337), the US Department of Energy (DE-FG03-02-ER40701), and the David and Lucile
Packard Foundation.
{}
\label{lastpage}
\end{document}